\documentclass[aps,showpacs,superscriptadress,preprint]{revtex4}
\usepackage{graphicx}

\begin{document}
\title{Scalar field quasinormal frequencies of Reissner-Nordstr$\ddot{\text{o}}$m black hole surrounded by  quintessence by using he continued fraction method }

\author{Chen Wu$^{1}$\footnote{Electronic address: wuchenoffd@gmail.com}} \affiliation{
\small 1. Shanghai Institute of Applied Physics, Chinese Academy of
Sciences, Shanghai 201800, China}

\begin{abstract}
We evaluate the  quasinormal modes of massless scalar field  around  Reissner-Nordstr$\ddot{\text{o}}$m black hole surrounded by a static and spherically symmetric quintessence by using the continued fraction method. The appropriate Frobenius series for  three special cases of the quintessence parameter $ \epsilon = -1/3, -2/3$ and $-1$ are derived successfully.  We show that  the variation of quasinormal frequencies with charge of the black hole and the quintessential parameters. The numerical results show that  quintessence field decreases oscillation frequencies of all angular momentum $l$ modes  and  increases the damping time of  $l>0$ modes.
\end{abstract}

 \maketitle

\section{Introduction}
Field perturbations around  black holes have been interest objects of discussions for decades. It has been well understood that the decay of fields in black hole background is dominated by some resonant quasinormal modes (QNMs) \cite{black hole}.  The QNMs are defined as the complex solutions to the perturbation wave equations under certain boundary conditions.  The study of the QNMs has gained considerable attention
  coming from AdS/CFT correspondence in string theory. The lowest quasinormal frequencies of black holes have a direct interpretation as dispersion relations of hydrodynamic excitations in the ultra-relativistic heavy ion collisions \cite{hydro}. Moreover, astrophysical interests in QNMs originated from their relevance in gravitational wave analysis. On September 14th, 2015, two advanced detectors of the Laser Interferometer Gravitational-wave Observatory (LIGO)  made the first direct measurement of gravitational waves  \cite{GW}.  The Advanced LIGO detectors observed a transient gravitational-wave signal determined to be the coalescence of two black holes, launching the era of gravitational-wave astronomy. All of these aspects motivated the  extensive numerical and analytical study of QNMs for different spacetime and different fields  around black holes \cite{some papers}. We refer the reader to the reviews \cite{review} where a lot of references to the recent research of QNMs can be found.

On the other hand, growing observational evidences indicating our universe are undergoing a stage of accelerated expansion, such as supernova of type Ia \cite{supernova}, the anisotropy of the cosmic microwave background radiation \cite{CMB}, and the large scale structure \cite{LSS}, indicate the presence of some mysterious form of repulsive energy called dark energy. In order to disclose the nature of dark energy, several models have been proposed.
The simplest option for this dark energy is the vaccum energy (the Einstein's cosmological constant) having  a constant equation of state with state parameter, $\epsilon = -1 $, but it needs extreme fine tuning to explain the observations. The second is the dynamical model of dark energy which is often realized by scalar field mechanism (quintessence \cite{quintessence}, phantom field \cite{phantom}, and the interaction model between dark energy and dark matter \cite{hejianhua}). Different models of dynamical dark energy are determined by the equation of state, namely the ratio of the pressure to the energy density.
 The quintessence can have an equation of state with $-1 \leq   \epsilon  \leq -1/3 $.  The cosmological constant with $\epsilon = -1$ can be considered as a special type of the quintessence.  If the evolution of the quintessence is slow enough and the kinetic energy density is less than the potential energy density, quintessence field can give rise to the negative pressure responsible for the  accelerated expansion.

 After Kiselev \cite{Kiselev} derived the exact solution of Einstein's field equations for quintessential matter surrounding a black hole, many authors considered the evolution of various fields around black holes surrounded by quintessence.  Chen and his coworkers  calculated the massless scalar field QNMs of the Schwarzschild black hole in the presence of quintessence  and observed that all angular momentum $l$ modes damps more rapidly due to the presence of quintessence \cite{chensongbai}. While  QNMs for other perturbation fields in quintessential Schwarzschild \cite{application1} and Reissner-Nordstr$\ddot{\text{o}}$m \cite{application2} were calculated. All these studies are based on the calculation of the QNMs  by using the third order WKB approximation method.

In this paper, we consider the  QNMs of massless scalar field perturbations around  Reissner-Nordstr$\ddot{\text{o}}$m black hole surrounded by   quintessence by using the continued fraction method. Since it is difficult to analyze the Frobenius series  for arbitrary values of the parameter $\epsilon$,  we take three special cases of the quintessence parameter, $\epsilon = -1/3, -2/3$ and $-1$,  so that the calculations become viable.
 The continued fraction method proposed by Leaver \cite{Leaver} is considered as the most accurate method to calculate the frequencies of perturbations, since this method has no intermediate approximation compared to other numerical methods. The key ingredient of this method is to cast the perturbation equation into a three-term recurrence relation, and from it we can get a continued fraction equation  to work out the QNMs.   Though the continued fraction method has been applied to more complex Kerr-Neumann de Sitter background \cite{KNds}, the  Frobenius series and the recurrence coefficients are not given explicitly. So it is interesting to implement the continued fraction method to investigate the scalar QNMs of RN-dS spacetime and our study fills this gap.  We derive the appropriate Frobenius series for three special cases of the quintessence parameter, $\epsilon$ = -1/3, -2/3 and -1 (this case is the RN-dS spacetime), and then obtain the  recurrent relation successfully. We find that the Leaver's method can also be  applied to the scalar perturbations of spherically symmetric black hole in de Sitter background. 

 The rest of the paper is organized as follows. In Sect. 2, we use the continued fraction method to study the scalar perturbations of Reissner-Nordstr$\ddot{\text{o}}$m black hole surrounded by quintessence and the results are presented. The conclusions are given in Sect. 3.

\section{The basic equations and numerical results }
The metric for the spacetime of  Reissner-Nordstr$\ddot{\text{o}}$m black hole surrounded by the spherically quintessential matte is given by \cite{Kiselev}
\begin{eqnarray}
ds^2 = -f(r) dt^2  + f(r)^{-1}dr^2  + r^2d\theta^2  + r^2\text{sin}^2\theta d\phi^2 ,
\end{eqnarray}
where
\begin{eqnarray}
f(r) =  1-\frac{2M}{r} + \frac{Q^2}{r^2} - \frac{c}{r^{3\epsilon+1} },
\end{eqnarray}
 $M$ is  the black hole mass, $Q$ is the charge of the black hole,  $\epsilon$ represents the state parameter of  quintessential and $c$ is the normalization factor which relates to the density of quintessence as, $\rho_q = \frac{-c}{2} \frac{3\epsilon}{r^{3(1-\epsilon)}}$. The Klein-Gordon equation describing the evolution of massless scalar perturbation field outside a charged black hole is given by
\begin{eqnarray}
   \frac{1}{\sqrt{-g}} \partial_\mu(g^{\mu\nu}\sqrt{-g} \partial_\nu\Phi) = 0.
\end{eqnarray}

Representing the scalar field in spherical harmonics,  one can reduce the wave equation  (3) in the  background to the Schr\"{o}dinger wave equation after separation of the angular variables
\begin{eqnarray}
\left( \frac{d^2}{dr^{*2}} + \omega^2 - V(r^*) \right) \Psi(r) = 0,
\end{eqnarray}
with the effective potential $V(r)$:
\begin{eqnarray}
V(r) = f(r) \left(  \frac{l(l+1)}{r^2} + \frac{2M}{r^3} -\frac{2Q^2}{r^4}+ \frac{c(3\epsilon+1)}{r^{3\epsilon+3}}  \right),
\end{eqnarray}
and
\begin{eqnarray}
   dr^* = \frac{dr}{f(r)}.
\end{eqnarray}

 It is difficult to analyze the above wave equation for arbitrary values of the parameter $\epsilon$ by means of the continued fraction method. For our study we take three special cases of the quintessence parameter, $\epsilon = -1/3,-2/3$
and -1, so that the calculations become viable. When $c=0$, the effective potential reduces to the one  for the Reissner-Nordstr$\ddot{\text{o}}$m spacetime.

\subsection{ the $\epsilon= -1/3$ case}

In this case, the black hole horizons are located at $r_{1+}$ and $r_{1-}$, corresponding to zeroes of the function $f(r)$ and we   can  write the tortoise coordinate as
\begin{eqnarray}
r_*=r + r_{1+} \textmd{ln}( r - r_{1+} ) + r_{1-} \textmd{ln}( r - r_{1-} ).
\end{eqnarray}

 Within the continued fraction method we can calculate the singular factor of the solution of the wave equation that satisfies in-going wave boundary condition at the event horizon and  out-going wave boundary condition at infinity, and  expand the remaining part into the Frobenius series that
are convergent in the  $ -\infty< r^* <+\infty$ region. The solution of the wave equation Eq. (4) is expanded as follows:
\begin{eqnarray}
\psi(r) = e^{i\omega r} (r - r_{1-})^{i\omega(r_{1+} + r_{1-})-1}  \sum_n a_n \left(  \frac{r-r_{1+}}{r-r_{1-}} \right)^{n+\delta},
\end{eqnarray}
where $\delta = \frac{-i\omega r^2_{1+} }{ r_{1+} - r_{1-} } $. The coefficients $a_n$ satisfy the three- term recurrence relations,
\begin{eqnarray}
\alpha_0 a_1 + \beta_0 a_0 &=& 0, \nonumber\\
\alpha_n a_{n+1} + \beta_n a_n + \gamma_n a_{n-1} &=& 0, \,\,\,\, n>0,
\end{eqnarray}
where
\begin{eqnarray}
\alpha_n = -(n+1)[ r_{1-}(n+1) +r_{1+}(-n+2i r_{1+} \omega -1)  ],
\end{eqnarray}
\begin{eqnarray}
 \beta_n &=& -r_{1+}[ l(l+1) + 2n^2-4ir_{1+}\omega (2n+1) +2n - 8r_{1+}^2\omega^2 + 1 ] \nonumber\\ &&
  + r_{1-} [l(l+1)+ 2n(n+1)+1 ] -2i(2n+1)r_{1-}r_{1+}\omega,
\end{eqnarray}
\begin{eqnarray}
\gamma_n = -[n-2i\omega(r_{1+} + r_{1-})] [n(r_{1-} - r_{1+}) + 2ir_{1+}^2\omega].
\end{eqnarray}

Since $r_{1+}$ is a regular singular point, convergence of the Frobenius series is automatically satisfied for $r_{1+} \leq r < \infty$.
Convergence  at $r= \infty$ implies convergence of the sum $\sum_n a_n$ and the coefficients $a_n$  satisfy the following infinite continued fraction equation:
\begin{eqnarray}
 \beta_{0}- \frac{\alpha_{0}\gamma_{1}}{ \beta_{1}- } \frac{\alpha_{1}\gamma_{2}}{\beta_2 -}\cdots =0.
\end{eqnarray}
Any fundamental QNM can be calculated by solving this  nonlinear equation  if the infinite continued fraction is truncated at some sufficiently large index.

\subsection{the $\epsilon= -2/3$ case}

 The black hole spacetime possesses a cosmological horizon at $r = r_{2c}$ except for the  horizons located at $r_{2+}$ and $r_{2-}$, with $r_{2c}>r_{2+}>r_{2-}$. Using these horizons  $f(r)$ can be written as
\begin{eqnarray}
 f(r)= \frac{c}{r^2}(r-r_{2+})(r-r_{2-})(r_{2c}-r).
\end{eqnarray}

Introducing the surface gravity  $\kappa_i= \frac{1}{2} \left| \frac{df}{dr} \right| _{r=r_i}$ associated with the horizon $r= r_i$, we can write the analytic form of the tortoise coordinate by calculating Eq. (6):
\begin{eqnarray}
r_* = \frac{1}{2\kappa_{2+}}  \textmd{ln}(\frac{r}{r_{2+}}-1) - \frac{1}{2\kappa_{2-}}  \textmd{ln}( \frac{r} { r_{2-} }-1 )
  + \frac{1}{2\kappa_{2c}}  \textmd{ln}( 1 - \frac{r} { r_{2c} } ) ,
\end{eqnarray}

 The appropriate Frobenius series  which satisfies the boundary conditions can be expanded  as
\begin{eqnarray}
\psi(r) &=& \left(\frac{1}{r} - \frac{1}{r_{2+} }\right)^{\frac{i\omega}{2k_{2+}}}  \left(\frac{1}{r} - \frac{1}{r_{2-}}\right)^{-\frac{i\omega}{2k_{2-}} + \frac{i\omega}{2k_{2+}} }  \left(\frac{1}{r} - \frac{1}{r_{2c}}\right)^{\frac{i\omega}{2k_{2c}}}  \nonumber\\ &&
\sum_n a_n \left(  \frac{1/r - 1/r_{2+}}{1/r - 1/r_{2-}} \frac{1/r_{2c} - 1/r_{2-}}{1/r_{2c} - 1/r_{2+}}\right)^n,
\end{eqnarray}
substituting this expansion into the wave equation, we  obtain a six-term recurrence relation for the coefficients $a_n$. The Gaussian eliminations allows us to determine the coefficients in the three-term recurrence relation and then scalar QNMs can be solved numerically using standard procedure. For simplicity, we do not list the six-term recurrence relation and the recurrence coefficients in detail.

\subsection{the $\epsilon= -1$ case}

This case corresponds to the Reissner-Nordstr$\ddot{\text{o}}$m-de Sitter (RNdS) spacetime. The zeroes of  $f(r)$ are corresponding to the event horizon at $ r r_{3+}$, the cauchy horizon at $ r_{3-}$, the cosmological horizon at $ r_{3c}$ and a negative zero at $ r_o = -r_{3-} - r_{3+}- r_{3c} $, respectively. Then  in terms of these horizons and the corresponding surface gravity  the tortoise coordinate is defined as
\begin{eqnarray}
r_* = \frac{1}{2\kappa_{3+}} \textmd{ln}(\frac{r}{r_{3+}}-1) - \frac{1}{2\kappa_{3-}} \textmd{ln}( \frac{r} { r_{3-} }-1 )
  - \frac{1}{2\kappa_{3c}} \textmd{ln}( 1 - \frac{r} { r_{3c} } )+ \frac{1}{2\kappa_{o}} \textmd{ln}( 1- \frac{r} { r_{o} } ).
\end{eqnarray}

The appropriate Frobenius series is found  to be
\begin{eqnarray}
\psi(r) &=& \left(\frac{1}{r} - \frac{1}{r_{3+} }\right)^{\frac{i\omega}{2k_{3+}}}  \left(\frac{1}{r} - \frac{1}{r_{3-}}\right)^{-\frac{i\omega}{2k_{3-}}  +  \frac{i\omega}{2k_{3+}}}  \left(\frac{1}{r} - \frac{1}{r_{3c}}\right)^{\frac{i\omega}{2k_{3c}}}  \left(\frac{1}{r} - \frac{1}{r_o }\right)^{-\frac{i\omega}{2k_{o}}}  \nonumber\\ &&
\sum_n a_n \left(  \frac{1/r - 1/r_{3+}}{1/r - 1/r_{3-}} \frac{1/r_{3c} - 1/r_{3-}}{1/r_{3c} - 1/r_{3+}}\right)^n,
\end{eqnarray}
substituting this series into the wave equation we obtain the following seven-term recursion relation for the coefficients $a_n$:
\begin{eqnarray}
 \beta_0 a_0 + \alpha_0 a_1 &=& 0, \nonumber\\
 \gamma_1 a_0 + \beta_1 a_1 + \alpha_1 a_2 &=& 0, \nonumber\\
\delta_2 a_0 + \gamma_2 a_1 + \beta_2 a_2 + \alpha_2 a_3 &=& 0, \nonumber\\
\eta_3 a_0 + \delta_3 a_1 + \gamma_3 a_2 + \beta_3 a_3+ \alpha_3 a_4 &=& 0, \nonumber\\
\theta_4 a_0 + \eta_4 a_1 + \delta_4 a_2 + \gamma_4 a_3 + \beta_4 a_4+ \alpha_4 a_5 &=& 0, \nonumber\\
\sigma_5 a_0 + \theta_5 a_1 + \eta_5 a_2 + \delta_5 a_3 + \gamma_5 a_4 + \beta_5 a_5+ \alpha_5 a_6 &=& 0, \nonumber\\
\sigma_n a_{n-5} + \theta_n a_{n-4} + \eta_n a_{n-3} + \delta_n a_{n-2} + \gamma_n a_{n-1} + \beta_n a_n+ \alpha_n a_{n+1} &=& 0,
\end{eqnarray}
where the recurrence coefficients are given in the Appendix. By repeating the Gauss elimination four times, a three-term recurrence relation can be derived.
Then the frequencies of the scalar perturbations are the solutions to the characteristic continued fraction equation.

\subsection{numerical results}

Now we report the frequencies of the  scalar perturbations in the Reissner-Nordstr$\ddot{\text{o}}$m black holes surrounded by quintessence. It is general experience that the fundamental QNMs will decay slowly and are relevant to the description of scalar field around the black hole. So we consider frequencies of fundamental modes for our study. We obtained the fundamental  QNMs  for different values of  the charge of the black hole Q,  the angular momentum index $l$, and  the quintessence parameter $\epsilon$. The results are summarized in Table 1.

The dependence of real and imaginary parts of fundamental QNMs on the charge of the black hole is plotted in Fig. 1 for fixed $l=2$, $c=0.005$ and for different values of $\epsilon$. In this figure, we put the normalization factor, $c=0.005$, so that the deviation of frequencies from the pure Reissner-Nordstr$\ddot{\text{o}}$m spacetime can be clearly understood. The figure shows that the QNMs for scalar field in charged black hole is influenced by quintessence. The magnitudes of real and imaginary parts of the $l=2$ modes is  lower in the presence of quintessential field. This implies that in the presence of quintessence, the $l=2$ modes for scalar field in Reissner-Nordstr$\ddot{\text{o}}$m black hole damp more slowly. It is found that $Re(\omega$) increases monotonically with the increase in $Q$ while the magnitude of $Im(\omega)$ first decreases, falling to a minimum and thereafter increases.   If we switch off the effect of quintessence, our results coincides with those obtained in Ref. \cite{RN QNM}.

Fig. 2 shows the explicit dependence of $Re(\omega)$ and $Im(\omega)$ with quintessential parameter $\epsilon$ and angular momentum $l$  fixed $ c= 0.005$ and $Q=0.5$. For  a fixed $l$, , $Re(\omega)$ increases slightly as the value of $\epsilon$ increases from $-1$ to $-1/3$. The magnitude of $Im(\omega)$ increases  as the value of $\epsilon$ increases for $l > 0$ modes while for $l=0$ mode the magnitude of $Im(\omega)$ first decreases, falling to a smaller value and thereafter increases again.

 \begin{table}[tbh]\centering
\caption{Fundamental quasinormal modes for  Reissner-Nordstr$\ddot{\text{o}}$m spacetime surrounded by  quintessence for different values of $Q$, $l$, $c$ and $\epsilon$.
 \vspace{0.3cm}} \label{table2}
\begin{tabular*}{16.5cm}{*{6}{c @{\extracolsep\fill}}}
\hline\hline $Q $ &$c$  &$l$ &$\omega (\epsilon = -1/3)$ &$\omega (\epsilon = -2/3)$ &$\omega (\epsilon = -1)$ \\
\hline
0.01  & 0.001 & 0 & 0.110347 -- 0.104792i & 0.109595 --0.104484i & 0.108712 -- 0.10470i \\
 &  &1 & 0.292648 -- 0.0975628i  & 0.291443 -- 0.0971104i & 0.288213 -- 0.096783i\\
  &  &2 & 0.483168 -- 0.0966625i  & 0.481362 -- 0.0961919i & 0.47655 --0.0956244i \\
% \hline
 0.01  & 0.005 & 0 & 0.109905 -- 0.104372i & 0.106146 -- 0.102819i  & 0.101025 -- 0.104226i\\
 &  &1 & 0.291476 -- 0.0971722i  & 0.285418 -- 0.0948987i & 0.268712 -- 0.0928568 i\\
  &  &2 & 0.481234 -- 0.0962755i  & 0.47214 -- 0.0939135i & 0.447269 -- 0.0907654i\\
 \hline
0.1  & 0.001 & 0 & 0.110538 -- 0.104833i & 0.109786 -- 0.104526i & 0.108907 -- 0.104743i\\
 &  &1 & 0.293138 -- 0.0976132i  & 0.291933 --0.0971613i  & 0.288712 -- 0.0968357i\\
  &  &2 & 0.483972 -- 0.0967143i  & 0.482165 -- 0.0962442i & 0.477367 -- 0.0956793i\\
% \hline
0.1  & 0.005 & 0 & 0.110095 -- 0.104413i  & 0.106336 --0.102863i & 0.101225 -- 0.104276i \\
 &  &1 & 0.291963 -- 0.0972222i & 0.285906 -- 0.0949515i & 0.269247 -- 0.0929212i\\
  &  &2 & 0.48203 -- 0.0963269i & 0.472939 -- 0.0939677i & 0.448139 -- 0.0908347i\\
 \hline

 0.5  & 0.001 & 0 & 0.115642 -- 0.105644  & 0.114891 -- 0.105349i & 0.114134 -- 0.105616i\\
 &  &1 & 0.30624 -- 0.0986983i  & 0.305043 -- 0.0982617i & 0.302049 -- 0.097982i\\
  &  &2 & 0.505437 -- 0.0978431i & 0.503643 -- 0.0973889i & 0.499189 -- 0.09689i\\
 %\hline
0.5  & 0.005 & 0 & 0.115156 -- 0.105219i  & 0.111400 -- 0.103728i & 0.106548-- 0.105305i\\
 &  &1 & 0.304954 -- 0.0982993i  & 0.298939 -- 0.0961056i & 0.283495 -- 0.0943666i\\
  &  &2 & 0.503315 -- 0.0974474i  & 0.494287 -- 0.0951677i & 0.471391- 0.0924144i\\
 \hline
0.9  & 0.001 & 0 & 0.132172 -- 0.102119i  & 0.131432 -- 0.10187i & 0.130244-- 0.102581i\\
 &  &1 & 0.352134 -- 0.0971095i  & 0.350982 -- 0.0967346i & 0.348699-- 0.096565i\\
  &  &2 & 0.581213 -- 0.0965454i & 0.579482 -- 0.0961557i & 0.56217-- 0.0958204i\\
% \hline
0.9  & 0.005 & 0 & 0.131536 -- 0.101798i  & 0.127835 -- 0.100546i & 0.123428-- 0.102363i\\
 &  &1 & 0.350353 -- 0.0967819i  & 0.344562 -- 0.0948978i  & 0.332872-- 0.093862i \\
  &  &2 & 0.578263 -- 0.0962164i & 0.569551 --0.0942599i &0.55224--0.0925138i \\
 \hline \hline
\end{tabular*}
\end{table}

\section{summary}
In this paper, we have calculated the  QNMs of massless scalar field  around  Reissner-Nordstr$\ddot{\text{o}}$m black hole surrounded by  spherically symmetric quintessence by using the continued fraction method.   In the extreme case of quintessence $\epsilon = -1$, i.e. the Reissner-Nordstr$\ddot{\text{o}}$m-de Sitter (RNdS) spacetime, we have derived a seven-term recurrence relation after Substituting appropriate Frobenius series encoding the QNMs' boundary conditions  into the radial equation of scalar field,  and then get the characteristic continued fraction equation by means of the the Gauss elimination successfully.
We found  that the magnitude of $Im(\omega)$ increases  as the value of $\epsilon$ increases for $l > 0$ modes while for $l=0$ mode the magnitude of $Im(\omega)$ first decreases, falling to a smaller value and thereafter increases again.

\begin{acknowledgments}
We thank Prof. Ru-Keng Su  for very helpful discussions. This work is supported partially by the Major State Basic Research Development Program in China (No. 2014CB845402).
\end{acknowledgments}

\appendix*
\section{}
In this section, the recurrence coefficients in Eqs. (19) are given explicitly as follow. The quintessence parameter $c$ is  taken as  $\lambda/3$ for convenience.
\begin{eqnarray}
\alpha_n &=& -(1+n) {r_{3+}}^2 ({r_{3+}}-{r_{3c}}) ({r_{3-}}-{r_{3c}}) (2 {r_{3+}}+{r_{3-}}+{r_{3c}})^2 \lambda  \nonumber\\&&
\left((1+n)({r_{3+}}-{r_{3-}}) ( {r_{3+}}-{r_{3c}}) (2 {r_{3+}}+{r_{3-}}+{r_{3c}}) \lambda +6 i {r_{3+}}^2 \omega \right)
\end{eqnarray}
\begin{eqnarray}
{\beta_n} &=& {r_{3+}} (-3 l {r_{3+}} ({r_{3+}}-{r_{3-}}) ({r_{3+}}-{r_{3c}})^2 (2 {r_{3+}}+{r_{3-}}+{r_{3c}})^2 \lambda -3 l^2 {r_{3+}} ({r_{3+}}-{r_{3-}}) ({r_{3+}}-{r_{3c}})^2  \nonumber\\&&
(2 {r_{3+}}+{r_{3-}}+{r_{3c}})^2 \lambda +    {4 (2+n+5 n^2) {r_{3+}}^8 \lambda ^2+(-1-2 n+2 n^2) {r_{3-}}^2 {r_{3c}}^3 ({r_{3-}}+{r_{3c}})^3 \lambda ^2-}   \nonumber\\&&
{{r_{3+}} {r_{3-}} {r_{3c}}^2 ({r_{3-}}+{r_{3c}})^2 ((-3-5 n+7 n^2) {r_{3-}}^2+(1+2 n-8 n^2) {r_{3-}} {r_{3c}}-2
(1+n)^2 {r_{3c}}^2) \lambda ^2-}  \nonumber\\&&
{r_{3+}}^2 {r_{3c}} ({r_{3-}}+{r_{3c}}) \lambda  (3+4 n-8 n^2) {r_{3-}}^4 \lambda +3 (-2-3 n+11 n^2) {r_{3-}}^3
{r_{3c}} \lambda +(1+2 n+4 n^2) \nonumber\\&&
 {r_{3c}}^4 \lambda +  {r_{3-}} {r_{3c}}^2 ((-2-4 n+n^2) {r_{3c}} \lambda +6 i \omega )+{r_{3-}}^2 {r_{3c}} ((-2-3 n+11 n^2) {r_{3c}} \lambda +12 i \omega ))- \nonumber\\&&
  {r_{3+}}^6 ((10+7 n+3 n^2) {r_{3-}}^2 \lambda ^2+2 {r_{3-}} \lambda  (n {r_{3c}} \lambda +37 n^2 {r_{3c}} \lambda -3 i \omega -78 i n \omega  ) +  ((2+n+5 n^2)  \nonumber\\&&
  {r_{3c}}^2 \lambda ^2+2 i (5+18 n) {r_{3c}} \lambda  \omega +12 \omega ^2))+
  {r_{3+}}^4 ((4+3 n-19 n^2)   {r_{3-}}^4 \lambda ^2+3 {r_{3-}}^3 \lambda  ((1+2 n+22 n^2)   \nonumber\\&&
    {r_{3c}} \lambda + 2 i (-1+2 n) \omega )+ {3 {r_{3-}}^2 {r_{3c}} \lambda  ((-1-2 n+14 n^2) {r_{3c}} \lambda -2 i (-5+18 n) \omega )+} \nonumber\\&&
{3 {r_{3c}}^2 ((1+n+3 n^2) {r_{3c}}^2 \lambda ^2+8 i (1+3 n) {r_{3c}} \lambda  \omega +24 \omega ^2)+} {r_{3-}} {r_{3c}} ((-7-10 n+42 n^2) {r_{3c}}^2 \lambda ^2 \nonumber\\&&
-24 i (-1+n) {r_{3c}} \lambda  \omega +72 \omega ^2))+  {r_{3+}}^5 (-(-1+n+35 n^2) {r_{3-}}^3 \lambda ^2+{r_{3-}}^2 \lambda ((21+22 n-8 n^2) {r_{3c}} \lambda \nonumber\\&&
 +12 i (-1+7 n) \omega )+  {3 {r_{3-}} ((5+5 n+3 n^2) {r_{3c}}^2 \lambda ^2-2 i (3+34 n) {r_{3c}} \lambda  \omega -12 \omega ^2)+}  \nonumber\\&&
{r_{3c}} ((11+16 n+38 n^2) {r_{3c}}^2 \lambda ^2-18 i (1+4 n) {r_{3c}} \lambda  \omega +72 \omega ^2))- {r_{3+}}^3 \lambda  ((-1-n+3 n^2) {r_{3-}}^5 \lambda  \nonumber\\&&
 +(9+10 n-44 n^2) {r_{3-}}^4 {r_{3c}} \lambda +{r_{3-}}
{r_{3c}}^3 ((9+13 n+13 n^2) {r_{3c}} \lambda -72 i n \omega)+  \nonumber\\&&
{2 {r_{3-}}^3 {r_{3c}} (2 (3+4 n+n^2) {r_{3c}} \lambda +3 i (-3+2 n) \omega )+3 {r_{3c}}^4 ((1+2 n+4 n^2)
{r_{3c}} \lambda -4 i (\omega +3 n \omega ))+}  \nonumber\\&&
 {r_{3-}}^2 {r_{3c}}^2 (4 (2+3 n+4 n^2) {r_{3c}} \lambda -3 i (\omega +4 n \omega )))+  {r_{3+}}^7 \lambda  ((-1+10 n^2) {r_{3-}} \lambda \nonumber\\&&
 -3 ((1+n+3 n^2) {r_{3c}} \lambda -i (\omega +6n \omega ))))
\end{eqnarray}
\begin{eqnarray}
{\gamma_n} &=&  {-\frac{1}{{r_{3-}}-{r_{3c}}}({r_{3+}}-{r_{3c}}) }
(-6 l {r_{3+}} ({r_{3+}}^2-{r_{3-}}^2) ({r_{3+}}-{r_{3c}})^2 (2 {r_{3+}}+{r_{3-}}+{r_{3c}})^2 \lambda -6 l^2 {r_{3+}} \nonumber\\&&
({r_{3+}}^2-{r_{3-}}^2) ({r_{3+}}-{r_{3c}})^2 (2 {r_{3+}}+{r_{3-}}+{r_{3c}})^2 \lambda +
20 {r_{3+}}^8 {r_{3-}} \lambda ^2-16 {r_{3+}}^7 {r_{3-}}^2 \lambda ^2-19 {r_{3+}}^6 {r_{3-}}^3 \lambda ^2 \nonumber\\&&
+7 {r_{3+}}^5 {r_{3-}}^4 \lambda^2  +7 {r_{3+}}^4 {r_{3-}}^5 \lambda ^2+{r_{3+}}^3 {r_{3-}}^6 \lambda ^2-12 {r_{3+}}^8 {r_{3c}} \lambda ^2-
16 {r_{3+}}^7 {r_{3-}} {r_{3c}} \lambda ^2+27 {r_{3+}}^6 {r_{3-}}^2 {r_{3c}} \lambda ^2 \nonumber\\&&
+28 {r_{3+}}^5 {r_{3-}}^3 {r_{3c}} \lambda ^2-14
{r_{3+}}^4 {r_{3-}}^4 {r_{3c}} \lambda ^2-12 {r_{3+}}^3 {r_{3-}}^5 {r_{3c}} \lambda ^2-{r_{3+}}^2 {r_{3-}}^6 {r_{3c}} \lambda ^2+  3 {r_{3+}}^6 {r_{3-}} {r_{3c}}^2 \lambda ^2  \nonumber\\&&
+6 {r_{3+}}^5 {r_{3-}}^2 {r_{3c}}^2 \lambda ^2-17 {r_{3+}}^4 {r_{3-}}^3 {r_{3c}}^2 \lambda
^2+3 {r_{3+}}^3 {r_{3-}}^4 {r_{3c}}^2 \lambda ^2+6 {r_{3+}}^2 {r_{3-}}^5 {r_{3c}}^2 \lambda ^2-{r_{3+}} {r_{3-}}^6 {r_{3c}}^2 \lambda ^2+ \nonumber\\&&
21 {r_{3+}}^6 {r_{3c}}^3 \lambda ^2-12 {r_{3+}}^5 {r_{3-}} {r_{3c}}^3 \lambda ^2-33 {r_{3+}}^4 {r_{3-}}^2 {r_{3c}}^3 \lambda ^2+16 {r_{3+}}^3
{r_{3-}}^3 {r_{3c}}^3 \lambda ^2+11 {r_{3+}}^2 {r_{3-}}^4 {r_{3c}}^3 \lambda ^2   \nonumber\\&&
-4 {r_{3+}} {r_{3-}}^5 {r_{3c}}^3 \lambda ^2 + {r_{3-}}^6 {r_{3c}}^3 \lambda ^2+3 {r_{3+}}^5 {r_{3c}}^4 \lambda ^2-14 {r_{3+}}^4 {r_{3-}} {r_{3c}}^4 \lambda ^2+15 {r_{3+}}^3 {r_{3-}}^2 {r_{3c}}^4 \lambda ^2  \nonumber\\&&
+3 {r_{3+}}^2 {r_{3-}}^3 {r_{3c}}^4 \lambda ^2-10 {r_{3+}} {r_{3-}}^4 {r_{3c}}^4 \lambda ^2+
3 {r_{3-}}^5 {r_{3c}}^4 \lambda ^2-9 {r_{3+}}^4 {r_{3c}}^5 \lambda ^2+12 {r_{3+}}^3 {r_{3-}} {r_{3c}}^5 \lambda ^2 \nonumber\\&&
+6 {r_{3+}}^2 {r_{3-}}^2 {r_{3c}}^5 \lambda ^2-12 {r_{3+}} {r_{3-}}^3 {r_{3c}}^5 \lambda ^2+3 {r_{3-}}^4 {r_{3c}}^5 \lambda ^2-
3 {r_{3+}}^3 {r_{3c}}^6 \lambda ^2+7 {r_{3+}}^2 {r_{3-}} {r_{3c}}^6 \lambda ^2   \nonumber\\&&
-5 {r_{3+}} {r_{3-}}^2 {r_{3c}}^6 \lambda ^2+{r_{3-}}^3
{r_{3c}}^6 \lambda ^2+ {(-1+n)^2 ({r_{3+}}-{r_{3-}}) ({r_{3+}}-{r_{3c}})^2 (2 {r_{3+}}+{r_{3-}}+{r_{3c}})^2 }  \nonumber\\&&
(4 {r_{3+}}^4+{r_{3+}}^3 (20 {r_{3-}}-6 {r_{3c}})-{r_{3-}}^2 {r_{3c}} ({r_{3-}}+{r_{3c}})+{r_{3+}} {r_{3-}} (3 {r_{3-}}^2-9
{r_{3-}} {r_{3c}}-8 {r_{3c}}^2)  \nonumber\\&&
+2 {r_{3+}}^2 (9 {r_{3-}}^2-7 {r_{3-}} {r_{3c}}-3 {r_{3c}}^2)) \lambda ^2+
24 i {r_{3+}}^8 \lambda  \omega +36 i {r_{3+}}^7 {r_{3-}} \lambda  \omega -12 i {r_{3+}}^6 {r_{3-}}^2 \lambda  \omega   \nonumber\\&&
-36 i {r_{3+}}^5
{r_{3-}}^3 \lambda  \omega -12 i {r_{3+}}^4 {r_{3-}}^4 \lambda  \omega -84 i {r_{3+}}^7 {r_{3c}} \lambda  \omega -
84 i {r_{3+}}^6 {r_{3-}} {r_{3c}} \lambda  \omega +60 i {r_{3+}}^5 {r_{3-}}^2 {r_{3c}} \lambda  \omega   \nonumber\\&&
+84 i {r_{3+}}^4 {r_{3-}}^3 {r_{3c}}
\lambda  \omega +24 i {r_{3+}}^3 {r_{3-}}^4 {r_{3c}} \lambda  \omega -48 i {r_{3+}}^6 {r_{3c}}^2 \lambda  \omega +
48 i {r_{3+}}^5 {r_{3-}} {r_{3c}}^2 \lambda  \omega \nonumber\\&&
+36 i {r_{3+}}^4 {r_{3-}}^2 {r_{3c}}^2 \lambda  \omega -24 i {r_{3+}}^3 {r_{3-}}^3
{r_{3c}}^2 \lambda  \omega -12 i {r_{3+}}^2 {r_{3-}}^4 {r_{3c}}^2 \lambda  \omega +72 i {r_{3+}}^5 {r_{3c}}^3 \lambda  \omega +  \nonumber\\&&
24 i {r_{3+}}^4 {r_{3-}} {r_{3c}}^3 \lambda  \omega -72 i {r_{3+}}^3 {r_{3-}}^2 {r_{3c}}^3 \lambda  \omega
-24 i {r_{3+}}^2 {r_{3-}}^3 {r_{3c}}^3 \lambda  \omega   +36 i {r_{3+}}^4 {r_{3c}}^4 \lambda  \omega \nonumber\\&&
 -24 i {r_{3+}}^3 {r_{3-}} {r_{3c}}^4 \lambda  \omega -
12 i {r_{3+}}^2 {r_{3-}}^2 {r_{3c}}^4 \lambda  \omega -72 {r_{3+}}^7 \omega ^2-180 {r_{3+}}^6 {r_{3-}} \omega ^2-108 {r_{3+}}^5 {r_{3-}}^2
\omega ^2 \nonumber\\&&
+180 {r_{3+}}^6 {r_{3c}} \omega ^2+360 {r_{3+}}^5 {r_{3-}} {r_{3c}} \omega ^2+
180 {r_{3+}}^4 {r_{3-}}^2 {r_{3c}} \omega ^2+180 {r_{3+}}^5 {r_{3c}}^2 \omega ^2+180 {r_{3+}}^4 {r_{3-}} {r_{3c}}^2 \omega ^2+  \nonumber\\&&
2 (-1+n) ({r_{3+}}-{r_{3c}}) (2 {r_{3+}}+{r_{3-}}+{r_{3c}}) \lambda  (2 {r_{3+}}^7 \lambda +{r_{3-}}^3 {r_{3c}}^2 ({r_{3-}}+{r_{3c}})^2 \lambda +{r_{3+}} {r_{3-}}^2 {r_{3c}}  \nonumber\\&&
 (-2 {r_{3-}}^3+{r_{3-}}^2
{r_{3c}}+4 {r_{3-}} {r_{3c}}^2+{r_{3c}}^3) \lambda +  {r_{3+}}^2 {r_{3-}} ({r_{3-}}^4 \lambda -5 {r_{3-}}^3 {r_{3c}} \lambda\nonumber\\&&
 -5 {r_{3c}}^4 \lambda +4 {r_{3-}} {r_{3c}}^2 (-2 {r_{3c}}
\lambda +3 i \omega )+{r_{3-}}^2 {r_{3c}} (-5 {r_{3c}} \lambda +12 i \omega ))+
{r_{3+}}^3 (2 {r_{3-}}^4 \lambda +2 {r_{3-}}^3 {r_{3c}} \lambda  \nonumber\\&&
 +3 {r_{3c}}^4 \lambda -4 {r_{3-}} {r_{3c}}^2 ({r_{3c}} \lambda +6 i \omega )-{r_{3-}}^2 {r_{3c}} ({r_{3c}} \lambda +12 i \omega ))+
{r_{3+}}^4 (-2 {r_{3-}}^3 \lambda +12 {r_{3-}}^2 ({r_{3c}} \lambda  \nonumber\\&&
+4 i \omega )+3 {r_{3c}}^2 (2 {r_{3c}} \lambda -11 i \omega )+{r_{3-}} {r_{3c}} (8 {r_{3c}} \lambda -57 i \omega ))+
{r_{3+}}^6 ({r_{3-}} \lambda -7 {r_{3c}} \lambda +18 i \omega ) \nonumber\\&& -{r_{3+}}^5 (4 {r_{3-}}^2 \lambda +{r_{3c}} (4 {r_{3c}}
\lambda +33 i \omega )-69 i {r_{3-}} \omega )))
\end{eqnarray}
\begin{eqnarray}
{\delta_n}  &=& \frac{1}{({r_{3-}}-{r_{3c}})^2}({r_{3+}}-{r_{3c}})^2   (-3 l ({r_{3+}}^3+3 {r_{3+}}^2 {r_{3-}}-3 {r_{3+}} {r_{3-}}^2-{r_{3-}}^3) ({r_{3+}}-{r_{3c}})^2  \nonumber\\&&
 (2 {r_{3+}}+{r_{3-}}+{r_{3c}})^2 \lambda -  3 l^2 ({r_{3+}}^3+3 {r_{3+}}^2 {r_{3-}}-3 {r_{3+}} {r_{3-}}^2-{r_{3-}}^3) ({r_{3+}}-{r_{3c}})^2  \nonumber\\&&
 (2 {r_{3+}}+{r_{3-}}+{r_{3c}})^2 \lambda + (-2+n)^2 ({r_{3+}}-{r_{3-}}) ({r_{3+}}-{r_{3c}})^2 (2 {r_{3+}}+{r_{3-}}+{r_{3c}})^2   \nonumber\\&&
({r_{3+}}^4+2 {r_{3+}}^3 (7 {r_{3-}}-2 {r_{3c}})+2 {r_{3+}} {r_{3-}} (7 {r_{3-}}^2-8 {r_{3-}} {r_{3c}}-6 {r_{3c}}^2)
+{r_{3-}}^2({r_{3-}}^2-4 {r_{3-}} {r_{3c}}-4 {r_{3c}}^2)+   \nonumber\\&&
 {r_{3+}}^2 (15 {r_{3-}}^2-8 {r_{3-}} {r_{3c}}-2 {r_{3c}}^2)) \lambda ^2+ (-2+n) ({r_{3+}}-{r_{3c}}) (2 {r_{3+}}+{r_{3-}}+{r_{3c}}) \lambda    \nonumber\\&&
(2 {r_{3+}}^7 \lambda +{r_{3-}}^3 {r_{3c}} (-{r_{3-}}^3+5 {r_{3-}}^2 {r_{3c}}+12 {r_{3-}} {r_{3c}}^2+6 {r_{3c}}^3)
\lambda +  {r_{3+}}^2 {r_{3-}} (5 {r_{3-}}^4 \lambda \nonumber\\&&
 -6 {r_{3c}}^4 \lambda -24 {r_{3-}} {r_{3c}}^2 ({r_{3c}} \lambda -i \omega )-{r_{3-}}^3
(11 {r_{3c}} \lambda +12 i \omega )+2 {r_{3-}}^2 {r_{3c}} (-7 {r_{3c}} \lambda +24 i \omega ))-  \nonumber\\&&
{r_{3+}}^5 (5 {r_{3-}}^2 \lambda +10 {r_{3-}} ({r_{3c}} \lambda -12 i \omega )+{r_{3c}} (7 {r_{3c}} \lambda +48 i \omega ))+ {r_{3+}}^4 (-12 {r_{3-}}^3 \lambda +12 {r_{3c}}^2 ({r_{3c}} \lambda -4 i \omega )   \nonumber\\&&
+3 {r_{3-}} {r_{3c}} (3 {r_{3c}} \lambda -56 i
\omega )+{r_{3-}}^2 (25 {r_{3c}} \lambda +192 i \omega ))+ {r_{3+}}^6 (7 {r_{3-}} \lambda -13 {r_{3c}} \lambda +12 i \omega ) \nonumber\\&&
{r_{3+}}^6 (7 {r_{3-}} \lambda -13 {r_{3c}} \lambda +12 i \omega )+{r_{3+}} {r_{3-}}^2 ({r_{3-}}^4 \lambda -10 {r_{3-}}^3 {r_{3c}}
\lambda -6 {r_{3c}}^4 \lambda -3 {r_{3-}}^2 {r_{3c}} ({r_{3c}} \lambda -8 i \omega )+\nonumber\\&&
24 i {r_{3-}} {r_{3c}}^2 \omega )+
 {r_{3+}}^3 ({r_{3-}}^4 \lambda +3 {r_{3c}}^4 \lambda +2 {r_{3-}}^3 (5 {r_{3c}} \lambda +12 i \omega )+{r_{3-}}^2 {r_{3c}}
(5 {r_{3c}} \lambda -48 i \omega )-\nonumber\\&&
-60 i {r_{3-}} {r_{3c}}^2 \omega )) 3 (-{r_{3-}}^3 {r_{3c}}^3 ({r_{3-}}+{r_{3c}})^3 \lambda ^2+{r_{3+}} {r_{3-}}^2 {r_{3c}}^2 ({r_{3-}}+{r_{3c}})^2 ({r_{3-}}^2+3{r_{3c}}^2) \lambda ^2+  \nonumber\\&&
{r_{3+}}^2 {r_{3-}} {r_{3c}} ({r_{3-}}+{r_{3c}}) \lambda  ({r_{3-}}^4 \lambda -3 {r_{3c}}^4 \lambda +{r_{3-}}^3 (3 {r_{3c}} \lambda
-2 i \omega )+4 {r_{3-}}^2 {r_{3c}} ({r_{3c}} \lambda +3 i \omega )+3 {r_{3-}} {r_{3c}}^2 \nonumber\\&&
({r_{3c}} \lambda +4 i \omega ))+ 4 {r_{3+}}^8 \lambda  ({r_{3-}} \lambda +{r_{3c}} \lambda -i \omega )-2 {r_{3+}}^7 (4 {r_{3-}}^2 \lambda ^2+{r_{3-}} \lambda  (4{r_{3c}} \lambda +9 i \omega )+(-13 i {r_{3c}} \lambda -\nonumber\\&&
-6 \omega ) \omega ) {r_{3+}}^3 {r_{3-}}^6 \lambda ^2+6 {r_{3-}} {r_{3c}}^5 \lambda ^2-{r_{3c}}^6 \lambda ^2+2 {r_{3-}}^5 \lambda  ({r_{3c}} \lambda
-i \omega )+{r_{3-}}^4 {r_{3c}} \lambda  (9 {r_{3c}} \lambda +22 i \omega )+ \nonumber\\&&
{r_{3-}}^3 {r_{3c}} (4 {r_{3c}}^2 \lambda ^2-i {r_{3c}} \lambda  \omega +12 \omega ^2)+3 {r_{3-}}^2 {r_{3c}}^2 (5
{r_{3c}}^2 \lambda ^2-8 i {r_{3c}} \lambda  \omega +16 \omega ^2))+\nonumber\\&&
{r_{3+}}^5 5 {r_{3-}}^4 \lambda ^2+2 {r_{3-}}^3 \lambda  (5 {r_{3c}} \lambda +8 i \omega )+2 {r_{3-}} {r_{3c}} (7 {r_{3c}}^2
\lambda ^2-2 i {r_{3c}} \lambda  \omega -96 \omega ^2)+\nonumber\\&&
 {r_{3-}}^2 (5 {r_{3c}}^2 \lambda ^2-i {r_{3c}} \lambda  \omega +27 \omega ^2)-{r_{3c}}^2 ({r_{3c}}^2 \lambda ^2+24
i {r_{3c}} \lambda  \omega +48 \omega ^2))+  {r_{3+}}^6 ({r_{3-}}^3 \lambda ^2 \nonumber\\&&
-{r_{3-}}^2 \lambda  ({r_{3c}} \lambda +8 i \omega )+{r_{3c}} (-7 {r_{3c}}^2 \lambda ^2+14
i {r_{3c}} \lambda  \omega -48 \omega ^2)+{r_{3-}} (-9 {r_{3c}}^2 \lambda ^2+46 i {r_{3c}} \lambda  \omega +84 \omega ^2))- \nonumber\\&&
{r_{3+}}^4 ({r_{3-}}^5 \lambda ^2+4 {r_{3-}}^4 \lambda  ({r_{3c}} \lambda -3 i \omega )-3 {r_{3c}}^4 \lambda  ({r_{3c}} \lambda -4
i \omega )+{r_{3-}}^3 (7 {r_{3c}}^2 \lambda ^2+44 i {r_{3c}} \lambda  \omega -36 \omega ^2)-\nonumber\\&&
 {r_{3-}} {r_{3c}}^2 ({r_{3c}}^2 \lambda ^2-3 i {r_{3c}} \lambda  \omega -18 \omega ^2)+{r_{3-}}^2 {r_{3c}}
(-{r_{3c}}^2 \lambda ^2+24 i {r_{3c}} \lambda  \omega +192 \omega ^2))))
\end{eqnarray}
\begin{eqnarray}
{\eta_n} &=&  \frac{1}{({r_{3-}}-{r_{3c}})^3}({r_{3+}}-{r_{3c}})^3
(6 l {r_{3-}} ({r_{3+}}^2-{r_{3-}}^2) ({r_{3+}}-{r_{3c}})^2 (2 {r_{3+}}+{r_{3-}}+{r_{3c}})^2 \lambda \nonumber\\&&
   +6 l^2 {r_{3-}}({r_{3+}}^2-{r_{3-}}^2) ({r_{3+}}-{r_{3c}})^2 (2 {r_{3+}}+{r_{3-}}+{r_{3c}})^2 \lambda - 4 (-2+n) {r_{3+}}^8 ((-10+3 n) {r_{3-}} \nonumber\\&&
-(-2+n) {r_{3c}}) \lambda ^2-
{{r_{3+}} {r_{3-}}^2 {r_{3c}} ({r_{3-}}+{r_{3c}}) \lambda  }  (-{r_{3c}} (11 {r_{3-}}^3+9 {r_{3-}}^2 {r_{3c}}+5 {r_{3-}} {r_{3c}}^2+7 {r_{3c}}^3) \lambda \nonumber\\&&
+2 (-3+n)^2 (4 {r_{3-}}^4-18{r_{3-}}^3 {r_{3c}}-4 {r_{3-}}^2 {r_{3c}}^2+11 {r_{3-}} {r_{3c}}^3+{r_{3c}}^4) \lambda +  (-3+n) \nonumber\\&&
 (-2 {r_{3-}}^4 \lambda +5 {r_{3c}}^4 \lambda +{r_{3-}}^3 (8 {r_{3c}} \lambda +6 i \omega )+2 {r_{3-}}^2 {r_{3c}} ({r_{3c}}
\lambda -15 i \omega )+{r_{3-}} {r_{3c}}^2 ({r_{3c}} \lambda -24 i \omega )))-  \nonumber\\&&
8 {r_{3+}}^7 \lambda  ((76-53 n+9 n^2) {r_{3-}}^2 \lambda -3 i (-2+n) {r_{3c}} \omega +{r_{3-}} (-2 (20-16 n+3
n^2) {r_{3c}} \lambda +  \nonumber\\&&
+3 i (-8+3 n) \omega )) {r_{3-}}^3 {r_{3c}}^2 ({r_{3-}}+{r_{3c}})^2 \lambda   (-3 {r_{3c}} ({r_{3-}}+{r_{3c}}) \lambda +2 (-3+n)^2 (2 {r_{3-}}^2-3 {r_{3-}} \lambda  \nonumber\\&&
 {r_{3c}}-3 {r_{3c}}^2) +(-3+n)
(-2 {r_{3-}}^2 \lambda +6 {r_{3-}} {r_{3c}} \lambda +6 {r_{3c}}^2 \lambda +6 i {r_{3-}} \omega ))+ {r_{3+}}^2 {r_{3-}} (2 (21-13 n+2 n^2) \nonumber\\&&
 {r_{3-}}^6 \lambda ^2+(52-40 n+7 n^2) {r_{3c}}^6 \lambda ^2+3 {r_{3-}}
{r_{3c}}^4 \lambda  ((71-38 n+5 n^2) {r_{3c}} \lambda -12 i (-4+n) \omega )+  \nonumber\\&&
{r_{3-}}^5 \lambda  ((-625+390 n-62 n^2) {r_{3c}} \lambda +6 i (-3+n) \omega )+2 {r_{3-}}^4 {r_{3c}} \lambda  ((161-102
n+16 n^2) {r_{3c}} \lambda -  \nonumber\\&&
6 i (-20+7 n) \omega )+ {r_{3-}}^3 {r_{3c}} ((928-580 n+91 n^2) {r_{3c}}^2 \lambda ^2  {r_{3-}}^2 {r_{3c}}^2 ((268-138 n+17 n^2) \nonumber\\&&
{r_{3c}}^2 \lambda ^2+24 i n {r_{3c}} \lambda  \omega -36 \omega ^2))+  {r_{3+}}^3 ((317-200 n+32 n^2) {r_{3-}}^6 \lambda ^2+(-2+n)^2 {r_{3c}}^6 \lambda ^2- \nonumber\\&&
4 {r_{3-}}^5 \lambda  (5 (67-43n+7 n^2) {r_{3c}} \lambda +3 i (5-2 n) \omega )+
12 {r_{3-}} {r_{3c}}^4 \lambda  ((-4+2 (-3+n)^2+n) {r_{3c}} \lambda  \nonumber\\&&
-2 i (-11+4 n) \omega )+  3 {r_{3-}}^2 {r_{3c}}^2 ((23-28 n+6 n^2) {r_{3c}}^2 \lambda ^2-8 i (-34+11 n) {r_{3c}} \lambda  \omega -108 \omega ^2)+ \nonumber\\&&
4 {r_{3-}}^3 {r_{3c}} ((-9-16 n+5 n^2) {r_{3c}}^2 \lambda ^2-18 i (-3+n) {r_{3c}} \lambda  \omega -90 \omega ^2)-
 {r_{3-}}^4 ((122-64 n+9 n^2) \nonumber\\&&
 {r_{3c}}^2 \lambda ^2+4 i (-1+2 n) {r_{3c}} \lambda  \omega -12 \omega ^2))-  {r_{3+}}^6 ((704-440 n+71 n^2) {r_{3-}}^3 \lambda ^2+{r_{3-}}^2 \lambda \nonumber\\&&
  ((-1160+788 n-131 n^2) {r_{3c}} \lambda +36 i (-26+9 n) \omega )+ {r_{3c}} (7 (-2+n)^2 {r_{3c}}^2 \lambda ^2-   \nonumber\\&&
12 i (-2+n) {r_{3c}} \lambda  \omega +36 \omega ^2)-  {r_{3-}} ((12-8 n+n^2) {r_{3c}}^2 \lambda ^2+8 i (-29+11 n) {r_{3c}} \lambda  \omega +36 \omega ^2))+ \nonumber\\&&
{r_{3+}}^5 ((308-230 n+41 n^2) {r_{3-}}^4 \lambda ^2+4 {r_{3-}}^3 \lambda  ((276-164 n+25 n^2) {r_{3c}}
\lambda -3 i (-91+30 n) \omega )- \nonumber\\&&
{r_{3c}}^2 ((-2+n)^2 {r_{3c}}^2 \lambda ^2+24 i (-2+n) {r_{3c}} \lambda  \omega +36 \omega ^2)-  4 {r_{3-}} {r_{3c}} ((110-87 n+16 n^2) {r_{3c}}^2 \lambda ^2  \nonumber\\&&
-6 i (-14+5 n) {r_{3c}} \lambda  \omega +90 \omega ^2)+
{r_{3-}}^2 (-2 (-28+n+2 n^2) {r_{3c}}^2 \lambda ^2+12 i (-135+46 n) {r_{3c}} \lambda  \omega \nonumber\\&&
 +324 \omega ^2))+
{r_{3+}}^4 ((725-472 n+78 n^2) {r_{3-}}^5 \lambda ^2+3 (-2+n) {r_{3c}}^4 \lambda  ((-2+n) {r_{3c}} \lambda -4 i \omega )+ \nonumber\\&&
{r_{3-}}^4 \lambda  ((-551+414 n-73 n^2) {r_{3c}} \lambda +12 i (24-7 n) \omega )- 2 {r_{3-}} {r_{3c}}^2 ((28-20 n+3 n^2) {r_{3c}}^2 \lambda ^2  \nonumber\\&&
+36 i (-8+3 n) {r_{3c}} \lambda  \omega +162 \omega ^2)+ {r_{3-}}^3 ((-389+244 n-36 n^2) {r_{3c}}^2 \lambda ^2 +12 i (-97+30 n) {r_{3c}} \lambda  \omega \nonumber\\&&
 +252 \omega ^2)- {r_{3-}}^2 {r_{3c}} ((909-562 n+86 n^2) {r_{3c}}^2 \lambda ^2-12 i (-23+6 n) {r_{3c}} \lambda  \omega +648 \omega
^2)))
\end{eqnarray}
\begin{eqnarray}
{\theta_n}&=& \frac{1}{({r_{3-}}-{r_{3c}})^4}{r_{3-}} ({r_{3+}}-{r_{3c}})^4
(-3 l ({r_{3+}}-{r_{3-}}) {r_{3-}} ({r_{3+}}-{r_{3c}})^2 (2 {r_{3+}}+{r_{3-}}+{r_{3c}})^2 \lambda \nonumber\\&&
-3 l^2 ({r_{3+}}-{r_{3-}}){r_{3-}} ({r_{3+}} -{r_{3c}})^2 (2 {r_{3+}}+{r_{3-}}+{r_{3c}})^2 \lambda +
 4 {r_{3+}}^7 ((43-23 n+3 n^2) {r_{3-}}  \nonumber\\&&
  +(-23+14 n-2 n^2) {r_{3c}}) \lambda ^2- {r_{3-}}^2 {r_{3c}}^2 ({r_{3-}}+{r_{3c}})^2 \lambda ((86-41 n+5 n^2) {r_{3-}}^2 \lambda  \nonumber\\&&
  +(-73+34 n-4 n^2) {r_{3c}}^2 \lambda +{r_{3-}} ((-73+34 n-4 n^2) {r_{3c}} \lambda +12 i (-4+n) \omega ))+ \nonumber\\&&
{r_{3+}} {r_{3-}} {r_{3c}} ({r_{3-}}+{r_{3c}}) \lambda
(2 (86-41 n+5 n^2) {r_{3-}}^4 \lambda -2 (-5+n)^2 {r_{3c}}^4 \lambda +{r_{3-}} {r_{3c}}^2 ((119-62 n+8 n^2) \nonumber\\&&
{r_{3c}} \lambda -24 i (-4+n) \omega )+  {r_{3-}}^3 ((-391+189 n-23 n^2) {r_{3c}} \lambda +24 i (-4+n) \omega )+{r_{3-}}^2 {r_{3c}} ((-2-3 \nonumber\\&&
(-4+n)^2+n) {r_{3c}} \lambda -48 i (-4+n) \omega ))+
4 {r_{3+}}^6 \lambda  ((-1+10 (-4+n)^2) {r_{3-}}^2 \lambda -6 i (-7+2 n) {r_{3c}}  \nonumber\\&&
\omega +{r_{3-}}(-(-5+9 (-4+n)^2+n) {r_{3c}} \lambda +3 i (-23+6 n) \omega ))+ {r_{3+}}^2 ((-86+41 n-5 n^2) {r_{3-}}^6 \lambda ^2 \nonumber\\&& {r_{3c}}^6 +(-23+14 n-2 n^2) \lambda ^2+{r_{3-}}^5 \lambda  ((735-358 n+44 n^2) {r_{3c}} \lambda -12 i (-4+n) \omega )-\nonumber\\&&
3 {r_{3-}} {r_{3c}}^4 \lambda  ((73-34 n+4 n^2) {r_{3c}} \lambda -4 i (-17+4 n) \omega )+{r_{3-}}^4 {r_{3c}} \lambda
 ((-37+18 n-2 n^2) {r_{3c}} \lambda \nonumber\\&&
 +6 i (-39+10 n) \omega )+ 2 {r_{3-}}^3 {r_{3c}} (-4 (68-33 n+4 n^2) {r_{3c}}^2 \lambda ^2+3 i (-17+4 n) {r_{3c}} \lambda  \omega +36 \omega ^2)+\nonumber\\&&
{r_{3-}}^2 {r_{3c}}^2 ((-226+99 n-11 n^2) {r_{3c}}^2 \lambda ^2+24 i (-9+2 n) {r_{3c}} \lambda  \omega +72 \omega
^2))+ {r_{3+}}^5    \nonumber\\&&
 ((215-95 n+11 n^2) {r_{3-}}^3 \lambda ^2+{r_{3-}}^2 \lambda  ((-945+460 n-56 n^2) {r_{3c}} \lambda +6 i (-119+30 n) \omega )+\nonumber\\&&
3 {r_{3-}} ((3-5 n+n^2) {r_{3c}}^2 \lambda ^2-2 i (-133+34 n) {r_{3c}} \lambda  \omega -36 \omega ^2)+ {r_{3c}} (7 (23-14 n+2 n^2) {r_{3c}}^2 \lambda^2 \nonumber\\&&
-12 i (-7+2 n) {r_{3c}} \lambda  \omega +72 \omega ^2))+   {r_{3+}}^3 ((-417+203 n-25 n^2) {r_{3-}}^5 \lambda ^2-3 {r_{3c}}^4 \lambda  ((23-14 n+2 n^2)  \nonumber\\&&
{r_{3c}} \lambda -4 i (-7+2 n) \omega )+ {r_{3-}}^4 \lambda  ((865-434 n+54 n^2) {r_{3c}} \lambda -6 i (-7+2 n) \omega )+
2 {r_{3-}}^3  \nonumber\\&&
(2 (55-26 n+3 n^2) {r_{3c}}^2 \lambda ^2-9 i (-9+2 n) {r_{3c}} \lambda  \omega -36 \omega ^2)+
4 {r_{3-}}^2 {r_{3c}} ((70-29 n+3 n^2) {r_{3c}}^2 \lambda ^2+\nonumber\\&&
6 i (-4+n) {r_{3c}} \lambda  \omega +72 \omega ^2)+ {r_{3-}} {r_{3c}}^2 ((1+13 n-3 n^2) {r_{3c}}^2 \lambda ^2+144 i (-4+n) {r_{3c}} \lambda  \omega +216 \omega ^2))+\nonumber\\&&
{r_{3+}}^4 ((-520+263 n-33 n^2) {r_{3-}}^4 \lambda ^2+{r_{3-}}^3 \lambda  ((-179+74 n-8 n^2) {r_{3c}} \lambda
+12 i (-33+8 n) \omega )+\nonumber\\&&
{r_{3-}}^2 ((113-47 n+5 n^2) {r_{3c}}^2 \lambda ^2-6 i (-155+38 n) {r_{3c}} \lambda  \omega -180 \omega ^2)+ {r_{3c}}^2 ((23-14 n+2 n^2) \nonumber\\&&
 {r_{3c}}^2 \lambda ^2+24 i (-7+2 n) {r_{3c}} \lambda  \omega +72 \omega ^2)+ {r_{3-}} {r_{3c}} ((643-312 n+38 n^2)  {r_{3c}}^2 \lambda ^2-6 i (-43+10 n) \nonumber\\&&
 {r_{3c}} \lambda  \omega +288 \omega ^2)))
\end{eqnarray}
\begin{eqnarray}
{\sigma_n} &=& \frac{1}{({r_{3-}}-{r_{3c}})^5}{r_{3-}}^2 ({r_{3+}}-{r_{3c}})^5 (-(-5+n)^2 ({r_{3+}}-{r_{3-}}) ({r_{3+}}-{r_{3c}})^2 ({r_{3-}}-{r_{3c}})  \nonumber\\&&
(2 {r_{3+}}+{r_{3-}}+{r_{3c}})^2 ({r_{3+}}+2 {r_{3-}}+{r_{3c}}) \lambda ^2-  6 i (-5+n) ({r_{3+}}-{r_{3c}}) (4 {r_{3+}}^4 ({r_{3-}}-{r_{3c}}) \nonumber\\&&
+{r_{3-}}^2 {r_{3c}} ({r_{3-}}+{r_{3c}})^2+{r_{3+}}^3 (6 {r_{3-}}^2-4 {r_{3-}} {r_{3c}}-6
{r_{3c}}^2)-2 {r_{3+}}^2 {r_{3c}} (-{r_{3-}}^2+2 {r_{3-}} {r_{3c}}+{r_{3c}}^2)+\nonumber\\&&
{r_{3+}} {r_{3-}}^2 (-{r_{3-}}^2+2 {r_{3-}} {r_{3c}}+3 {r_{3c}}^2)) \lambda  \omega +36 {r_{3+}}^2 ({r_{3+}}^2
({r_{3-}}-{r_{3c}})-{r_{3-}} {r_{3c}} ({r_{3-}}+{r_{3c}})\nonumber\\&&
+{r_{3+}} ({r_{3-}}^2-2 {r_{3-}} {r_{3c}}-{r_{3c}}^2)) \omega^2)
\end{eqnarray}

\begin{figure}[tbp]
\includegraphics[width=13cm,height=19cm]{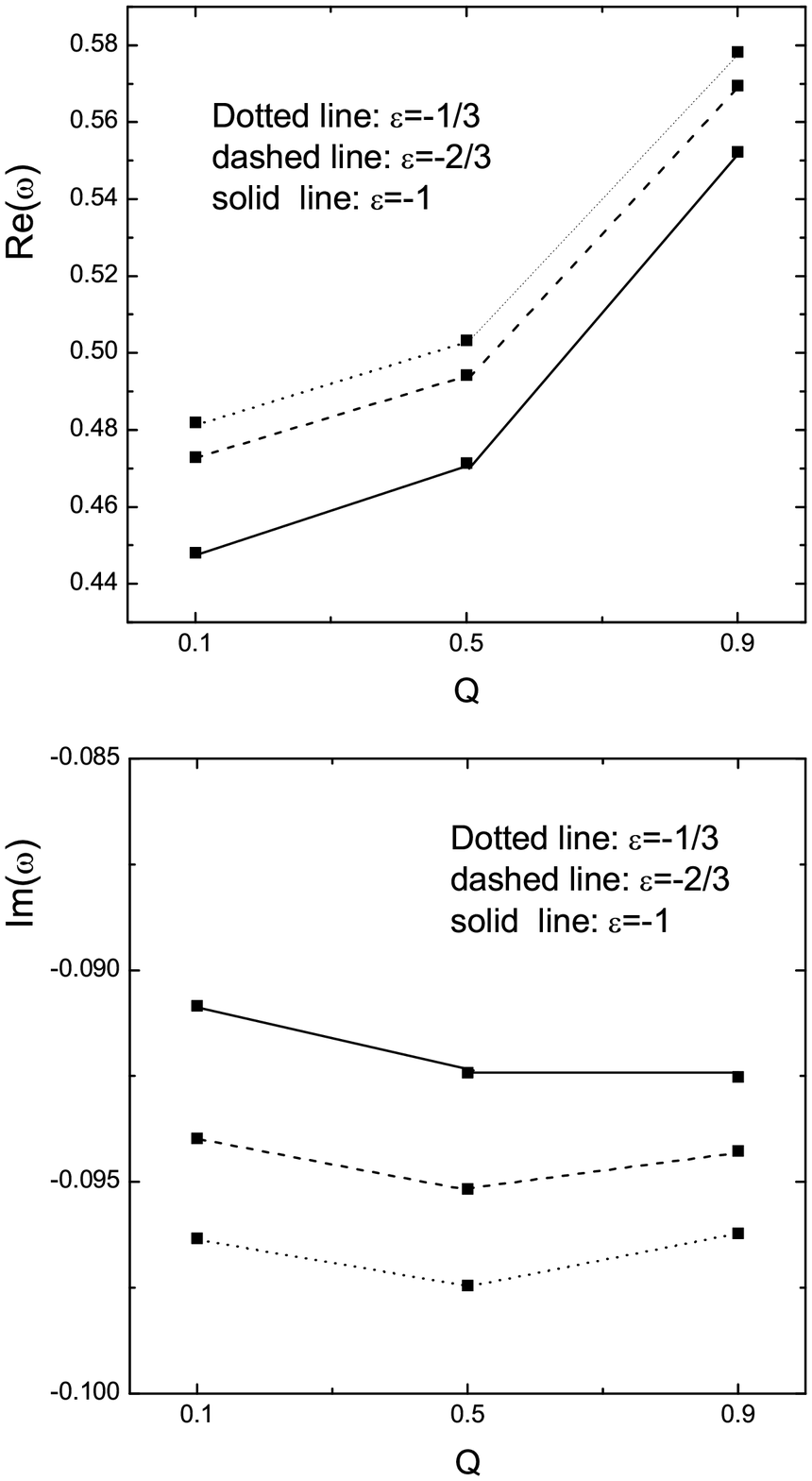}
\caption{ $Re(\omega)$ and $Im(\omega)$ for scalar field in Reissner-Nordstr$\ddot{\text{o}}$m spacetime surrounded by  quintessence as a function of $Q$ for $l = 2$ and for different values of $\epsilon$ with $c = 0.005$.    }
\end{figure}

\begin{figure}[tbp]
\includegraphics[width=13cm,height=19cm]{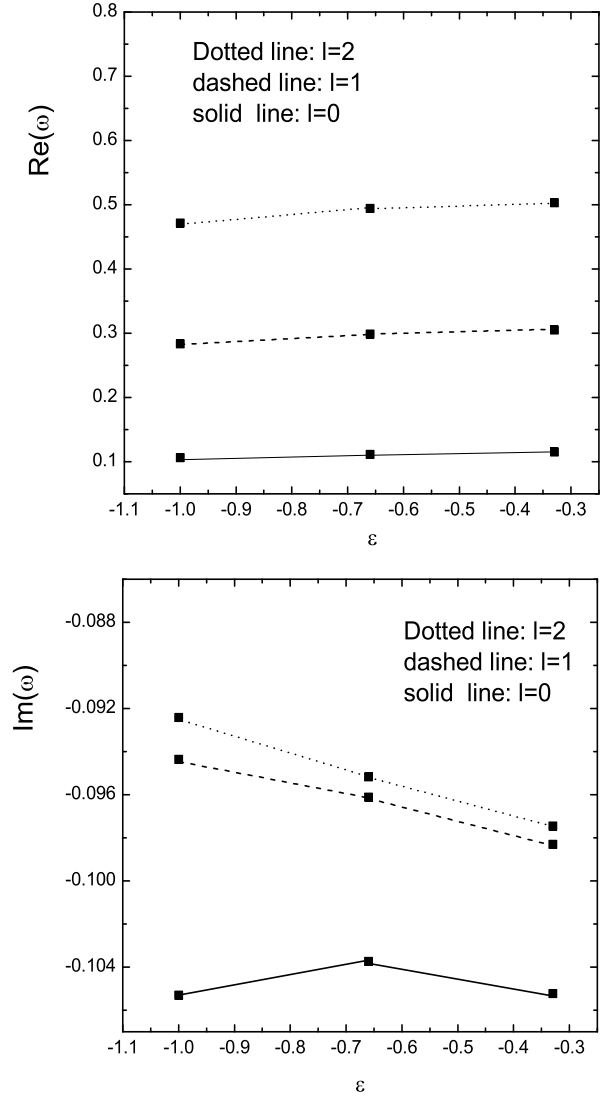}
\caption{ $Re(\omega)$ and $Im(\omega)$ for scalar field in Reissner-Nordstr$\ddot{\text{o}}$m spacetime surrounded by  quintessence  as a function of $\epsilon$ for $Q = 0.5$ and $c = 0.005$ and for different values of $l$ . }
\end{figure}

\end{document}